# THE STUDY ABOUT THE ANALYSIS OF RESPONSIVENESS PAIR CLUSTERING TO SOCIAL NETWORK BIPARTITE GRAPH

Akira Otsuki [1] and Masayoshi Kawamura[2]

[1] Tokyo Institute of Technology, Tokyo, Japan

`cecil2005@hotmail.co.jp`

[2]MK future software, Ibaraki, Japan

`kawamura.masa@nifty.com`

## ABSTRACT

*In this study, regional (cities, towns and villages) data and tweet data are obtained from Twitter, and extract information of "purchase information (Where and what bought)" from the tweet data by morphological analysis and rule-based dependency analysis. Then, the "The regional information" and the "The information of purchase history (Where and what bought information)" are captured as bipartite graph, and Responsiveness Pair Clustering analysis (a clustering using correspondence analysis as similarity measure) is conducted. In this study, since it was found to be difficult to analyze a network such as bipartite graph having limitations in links by using modularity Q, responsiveness is used instead of modularity Q as similarity measure. As a result of this analysis, "regional information cluster" which refers to similar " The information of purchase history" nodes group is generated. Finally, similar regions are visualized by mapping the regional information cluster on the map. This visualization system is expected to contribute as an analytical tool for customers' purchasing behaviour and so on.*

## 1. INTRODUCTION

"Big data" is not just meaning the size of the capacity simply. For example, it is real-time data or nonstructural data, like the SNS (Social Networking Service) data. Big data is being used in various fields as environmental biology [1], physics simulation [2], seismology, meteorology, economics, and management information science.

Foreign countries are making efforts aggressive towards big data utilization already according to the data of Ministry of Internal Affairs and Communications [3]. OSTP (Office of Science and Technology Policy) in the US government released the "Big Data Research and Development Initiative" at March 2013. Other hand, FI-PPP (Future Internet Public-Private Partnership) program is being implemented by EU at 2011. FI-PPP is a Public-Private-Partnership programme for Internet-enabled innovation. In this manner, the study of Big Data is being implemented actively worldwide as a recent trend. Therefore, it is conceivable that the study of Big Data is very important.

We [4-6] did propose many Big Data analysis methods thus far. These are the methods based on Bibliometrics and clustering (Newman Method). We will do Study about the Analysis of "Responsiveness Pair Clustering" of Bipartite Graph of Regional / Purchase Big Data in this paper. Concretely, first will get the Regional (Ex: cities, towns and villages) data and purchase data from Twitter. Next, will do "Responsiveness Pair Clustering Analysis" about bipartite graph of regional and purchase data. Although former clustering method used the structure of the link as the similarity measure of clustering, the "Responsiveness Pair Clustering Analysis" is the analysis method using the responsiveness [7-9] of the data as the similarity measure of clustering. The cluster that purchasing behaviour is the same is created after the result of this analysis. Then similar area (cities, towns and villages) will be visualization by mapping these clusters on the map. This visualization system is expected to contribute as an analytical tool for customers' purchasing behaviour and so on.

## 2. RELATED WORK AND BASIC TECHNOLOGY OF BIG DATA ANALYSIS

### 2.1. Bibliometrics

Bibliometrics is the analysis method of citation relationship proposed by Garfield [10-13]. It analyse citation relation as a target of Journal papers. There are three techniques in the citation relation analysis as shown in ① - ③ following.

① **Direct Citation**

As shown in Fig1, Papers A and B are cited in Paper C, In this case, direct citation deems that there are links between Papers A/B and Paper C and further links between Paper C. As a result, there are three nodes and two links in the network. When direct citation is used, a certain paper is deemed to have links with all papers that cite the pertinent paper.

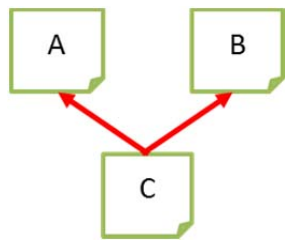

Figure 1. Direct Citation

② **Co-Citation**

This was proposed by Small [14]. As shown in Fig2, both Paper A and Paper B are cited in Paper C. In this case, co-citation deems that there is a link between Paper A and Paper B; thus, there are two nodes and one link in the network. For pairs of papers in which co-citation was used, i.e., all papers contained in the list of cited literature of a certain paper, there is a link between the paired papers.

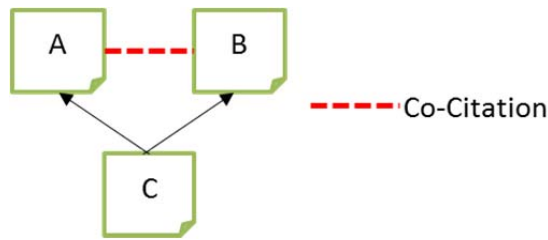

Figure 2. Co-Citation

③ **Bibliographic coupling**

It is a technique proposed by Kessler [15]. As shown in Fig3, both Paper D and Paper E cited Paper C. In this case, this technique deems that there is a link between Paper D and Paper E; thus, there are two nodes and one link in the network. When bibliographic coupling is used for pairs of papers that cite a certain paper, it is deemed that there is a link between the paired papers.

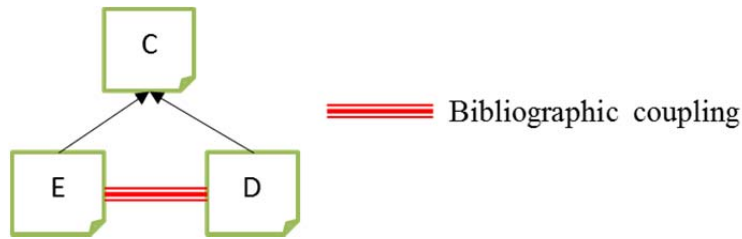

Figure 3.  Bibliographic coupling

## 2.2. Clustering method based on the Modularity $Q$

Clustering method is based on the graph theory. Clustering is a technique used to divide a large volume of data like academic papers. According to common features by clustering can simplify the overall structure of complex data and understand it more directly and thoroughly. There are a few techniques at Clustering Methods as shown in A) - C) below.

**A) Newman Method [16-18]**

   This technique is the clustering technique proposed by M. E. J. Newman. This technique does the clustering by optimizing the Modularity $Q$.

**B) GN Method [19]**

   This technique proposed by M.E.J.Newman and M.Girvan. This technique does the clustering by using betweenness of the edge.

**C) CNM Method [20]**

   This technique proposed by Aaron Clauset, M.E.J.Newman and Cristopher Moore. This technique is the faster technique than Newman method.

Fig4 is the citation map by Newman Method. Fig4 is the example that did divide a large volume of data like academic papers about "Data Mining". In this way, will be able to simplify the overall structure of complex data and understand it more directly and thoroughly by using clustering method.

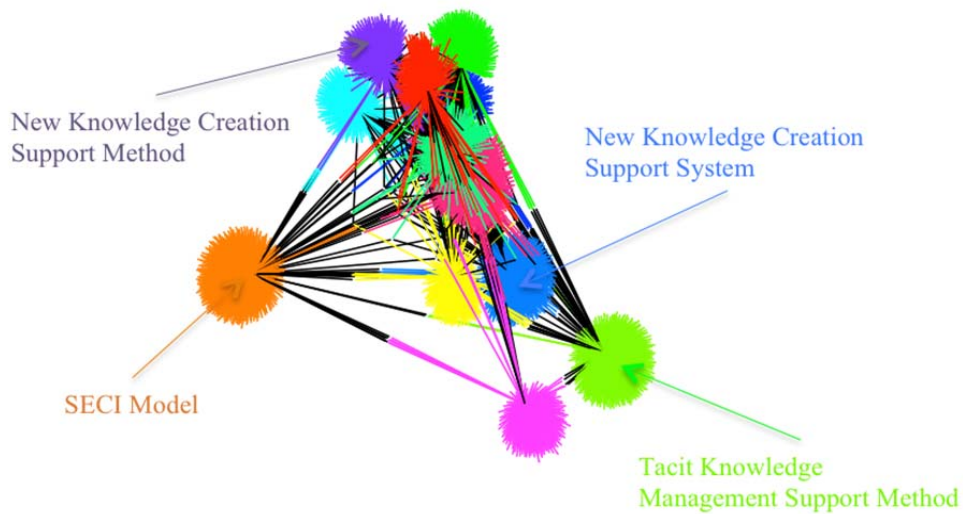

Figure 4.  Example of the clustering (using Newman Method)

## 2.3. The Problem of the Modularity $Q$ in the Bipartite Graph

The modularity $Q$ can handle the network there is no restriction on the link as shown in Table1. For example, "Akira, O.2000" has appeared in two both the column at the Table1. This means that there is no restriction on the link. But modularity $Q$ can't handle the network (Bipartite Graph) there is restriction on the link like as shown in Table2. For example, there is no the data that appear in both two columns in the Table 2. This means that there is restriction on the link.

Table 1.  There is no restriction on the link of network

| Paper Name | Cited Papers Name |
|---|---|
| **Akira, O.2000** | Author, A2013 |
| **Akira, O.2000** | Author, B2011 |
| **Akira, O.2000** | Author, C2012 |
| Masayoshi, K.1995 | Author, D2012 |
| Masayoshi, K.1995 | **Akira, O.2000** |
| Masayoshi, K.1995 | Author, E2012 |
| Author, E2012 | Author, F2012 |

Table 2.  There is restriction on the link of network

| Purchases Item | Purchases Place |
|---|---|
| Electrical appliances | Electrical appliance store |
| Electrical appliances | Electrical appliance store |
| Dress | Department store |
| Accessories | Department store |
| Cake | Supermarket |

## 2.4. Related Work

There are many studies that apply modularity $Q$ to bipartite graph. $Q^B$ is the bipartite modularity proposed by Barber [21]. $Q^B$ is shown as formula (1):

$$Q^B = \frac{1}{2m}\sum_{i=1}^{n}\sum_{j=1}^{n}(A_{ij} - P_{ij})\delta(c(v_i), c(v_j))$$

(1)

$P_{ij}$ shows the probability there are the link to the $V_i$ and $V_j$ on the random bipartite graph. $P_{ij}$ is shown as formula (2):

$$P_{ij} = \begin{cases} \dfrac{k_i k_j}{m} & If\ v_i v_j\ are\ belong\ to\ different\ a\ subset \\ 0 & Other \end{cases}$$

(2)

$A_{ij}$ shows the element of adjacency matrix. $A_{ij}$ is shown as formula (3):

$$A_{ij} = \begin{cases} 1 \ if \ (i,j) \in E \\ 0 \quad Other \end{cases}$$

(3)

Barber proposed the method of community divide at the maximum $Q$.

Takeshi, M. [22] did proposed bipartite Modularity $Q^M$ by giving the correspondence relation to the difference community sets.

$$Q^M = \sum_i (e_{ij} - a_i a_j), j = \begin{array}{c} argmax(e_{ik}) \\ k \end{array}$$

(4)

$Q^M$ is evaluate the link density the most corresponds community. But it's mean that $Q^M$ can only evaluate about the one community.

Kazunari, I. [23] proposed the bipartite modularity $D_{ij}^{Hf}$ in order to address this issue. $D_{ij}^{Hf}$ is shown as formula (5).

$$D_{ij}^{Hf} = \sum_{k=1}^{n_t} |r_{ik} - r_{jk}|, \ IBRP_k = \frac{1}{(n_t)^2} \sum_{i,j \in F}^{n_t} D_{ij}^{Hf}$$

(5)

$D_{ij}^{Hf}$ is a partitioning algorithm known as the Weakest Pair (WP) algorithm. This separates the weakest pairs of bloggers and webpages, respectively, using co-citation information.

Finally, Keiu, H. [24] proposed a new bipartite modularity $Q^H$ which is a measure to evaluate community structure considering correspondence of the community relation quantitatively.

$$Q^H = \sum_{i=1}^{|C_A|} \sum_{j=1}^{|C_B|} a_{ij}(e_{ij} - a_i a_j)$$

(6)

He showed $C = C_A \cup C_B$ as a bipartite graph community when each part communities are $C_A$ and $C_B$. $(e_{ij}-a_i a_j)$ compute the difference between the expected value of link density and the density of links between communities, like Modularity Q. But $Q^H$ can evaluate the relationship between communities, but can't evaluate relationship between nodes in the communities.

## 3. PROPOSED METHOD OF THIS STUDY

This study will propose the analysis method of responsiveness pair clustering to social network bipartite graph. Then, will do implementation the social network bipartite graph visualization system as a target to twitter data. This system is expected as a marketing system about customer purchase behaviour.

### 3.1. Acquisition of the Target Data

We used our own script, to which the twitter API is applied, to obtain information from tweeted comments about what commercial items consumers purchased during the Christmas period and where, in order to use it as analysis data. Specifically, we acquired tweets and position (latitude/longitude) information on from December 22 to 25, 2012. About 750 pieces of information were obtained. Fig5 shows an example of formatted information.

(1) Tue, 22 Dec 2012,

(2) twitter_User_ID,

(3) I bought the clothes by **department store. (@ **Department store w/7 others),

(4) [35.628227, 139.738712]

Figure 5. Example of obtained information's (tweeted comments and position information)

## 3.2. Morphological Analysis and Dependency Parsing

We'll explain analysis method about obtained information using morphological analysis and dependency parsing in this section.

First, Geocoding was applied to extract the names of cities, towns and villages (position data) from the latitude and longitude shown in (4) of Fig5. We then extracted information about what items were purchased by consumers and where from their tweets {(3) of fig5}. As for the shopping location, the part "@ so-and-so department" in (3) was automatically extracted. A morphological analysis was performed on the information of (3) using ChaSen [25], and then a rule-based Dependency Parsing [26] was done to extract information of purchased items. As a result, a "clothes" was extracted in the case of Fig5. We manually extracted these information if had not been extracted automatically. Dependency parsing is the method of analyse relation words about each words as shown in Fig6.

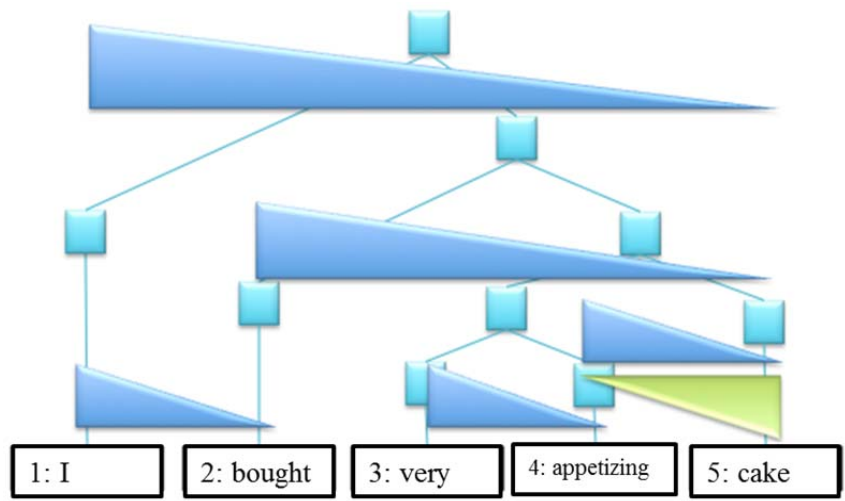

| 1: I | 2: bought | 3: very | 4: appetizing | 5: cake |

Figure 6. Dependency parsing

We have constructed the rule-based for extracting the subject from the verb, and then it applied to dependency parsing. The analysis target data for "Responsiveness Pair Clustering analysis (next section)" are created after the above analysis (Geocoding, morphological analysis and rule-based dependency parsing) as a bipartite graph as shown in the table3. The left column of table3 shows the purchase information and the right column shows cities, towns and villages (position information). We will propose the method of "Responsiveness Pair Clustering analysis" using the bipartite graph data (table3) in the next section.

Table 3.  The analysis target data for "Responsiveness Pair Clustering analysis (next section)"

| The information of Purchase _ Purchase place | City |
|---|---|
| Hair dryer _ Home electronics retailer | Shinjuku-ku, Tokyo, Japan |
| Refrigerator _ Home electronics retailer | Asaka-City, Saitama, Japan |
| Clothing _ Department store | Osaka-shi, Osaka, Japan |
| Cake _ Department store | Ichikawa-city,Chiba, Japan |
| Cake _ Supermarket | Oita-city, Oita, Japan |
| Cake _ Department store | Adachi-ku, Tokyo, Japan |
| Desk _ Supermarket | Akita-city, Akita, Japan |

### 3.3. Responsiveness Pair Clustering analysis

It is difficult to using bipartite graph at the Modularity $Q$ as shown in section 2.3. Therefore, this study conduct the hierarchical clustering using responsiveness (the similarity measure) between 2 parts of the dataset (table3), in place of the modularity.

First of all, we consider the responsiveness to be used as the similarity measure for clustering. Concretely, we set the similarity measure using MCA (Multiple Correspondence Analysis) by reference to the method used by Vanables, W.N. [27], which is generally used in the statistical software "R". Then, the binary cross tabulation of (m×n) should be set as an initial matrix for MCA, which can be expressed as the below formula.

$$\underset{m\times n}{F} = \left( f_{ij} \right) \ \left( f_{ij} \geq 0, i \in I, j \in J \right) \tag{7}$$

$I$ and $J$ represent the set of alternatives for the items of each row and column as expressed below.

$$I=\{1,2,\dots,m\}, \quad J=\{1,2,\dots,n\} \tag{8}$$

The concept for profile is assumed as the patterns of relative ratio of the rows or columns of the cross tabulations as indicated as shown in the below (9) and (10).

$$N_I = \left\{ q_{ij} = \frac{p_{ij}}{p_{i+}} | i \in I, j \in J \right\} \quad \text{(The profile of row)} \tag{9}$$

$$N_J = \left\{ q_{ij}^* = \frac{p_{ij}}{p_{+j}} | i \in I, j \in J \right\} \quad \text{(The profile of column)} \tag{10}$$

Secondly, we considering about MCA. If the variables are dichotomized, the binary cross tabulation should be set as an initial matrix, then that matrix should be made firstly to apply $x_{ij}$, then the below (11) and (12) as its elements.

$$x_{ij} = \frac{p_{ij}}{p_{i+}\sqrt{p_{+j}}} - \sqrt{p_{+j}} = \frac{q_{ij}}{\sqrt{p_{+j}}} - \sqrt{p_{+j}} \tag{11}$$

$$x_{ij}^* = \frac{p_{ij}}{p_{j+}\sqrt{p_{i+}}} - \sqrt{p_{i+}} = \frac{q_{ij}^*}{\sqrt{p_{i+}}} - \sqrt{p_{i+}} \tag{12}$$

Subsequently, the elements of $x_{ij}$ can be expressed as below, and it is the basic matrix for MCA.

$$\underset{m\times n}{X} = \left( x_{ij} \right) (i \in I, j \in J) \tag{13}$$

Thirdly, we think about the component score of responsiveness pair analysis. Formula (14) and (15) are the component scores of purchase information and regional information. The component score of the k-th for the purchase node $i$ is as the below formula.

$$z_{ik} = (i \in I, k = 1, 2, \cdots K) \tag{14}$$

The component score of the k-th for the regional node j is as the below formula.

$$z_{jk} = (j \in J, k = 1, 2, \cdots K) \tag{15}$$

Then, The relationship of probability matrix and the component scores of $I$ and $J$ are shown in table4.

Table 4.  The relationship of probability matrix and the component scores of $I$ and $J$

| | | 1 | 2 | ... | J | ... | n | Row sum |
|---|---|---|---|---|---|---|---|---|
| | 1 | $f_{11}$ | $f_{12}$ | ... | $f_{1j}$ | ... | $f_{1n}$ | $f_{1+}$ |
| | 2 | $f_{21}$ | $f_{22}$ | ... | $f_{2j}$ | ... | $f_{2n}$ | $f_{2+}$ |
| | $\vdots$ | $\vdots$ | $\vdots$ | $\vdots$ | $\vdots$ | $\vdots$ | $\vdots$ | $\vdots$ |
| $I$ | I | $f_{i1}$ | $f_{i2}$ | ... | $f_{ij}$ | ... | $f_{in}$ | $f_{i+}$ |
| | $\vdots$ | $\vdots$ | $\vdots$ | | $\vdots$ | | $\vdots$ | $\vdots$ |
| | m | $f_{m1}$ | $f_{m2}$ | ... | $f_{mj}$ | ... | $f_{mn}$ | $f_{m+}$ |
| | Column sum | $f_{+1}$ | $f_{+2}$ | ... | $f_{+j}$ | ... | $f_{+n}$ | $f_{++}$ |

The component scores ($z_{ik}$, $z_{jk}$) calculated as above (14) and (15) are used as the coordinates of matrix for hierarchical clustering.

Next, we will consider the hierarchical clustering. Generally in case of 2 variables, the hierarchical clustering generates the clusters by merging those whose Euclidian distances are shorter by calculating it between $i$ and $j$. Then, 2 clusters that indicated the shortest distances between each other are sequentially merged, so that it can obtain the hierarchical structure by repeated integration of all subjects into the final one cluster. In case of 2 variables, the Euclidian distance between the subjects i and j can be expressed as below, if the coordinate between i and j is set as ($x_{i1}$, $x_{j1}$).

$$d_{ij} = \sqrt{(x_{i1} - x_{j1})^2 + (x_{i2} - x_{j2})^2} \tag{16}$$

Also for multivariate cases, it should be defined as below by extending the formula (16).

$$d_{ij} = \sqrt{\sum_{k=1}^{n}(x_{ik} - x_{jk})^2}$$

(17)

By applying above $z_{ik}$ and $z_{jk}$ to the multivariate coordinates of the formula (17), we can calculate the Euclidian distance by using the correspondence relation as the similarity measure. This should be expressed as the next formula.

$$d_{ij} = \sqrt{\sum_{s=1}^{n}(z_{ik_s} - z_{jk_s})^2}$$

(18)

Next, there are a many methods of measuring the inter-cluster as shown in the following.

● **Nearest neighbor method**

$$d(C_1, C_2) = \min_{x_1 \in C_1, x_2 \in C_2} d(x_1, x_2)$$

(19)

● **Furthest neighbor method**

$$d(C_1, C_2) = \max_{x_1 \in C_1, x_2 \in C_2} d(x_1, x_2)$$

(20)

● **Group average method**

$$d(C_1, C_2) = \frac{1}{|C_1||C_2|} \sum_{x_1 \in C_1} \sum_{x_2 \in C_2} d(x_1, x_2)$$

(21)

● **Ward method**

$$d(C_1, C_2) = E(C_1 \cup C_2) - E(C_1) - E(C_2)$$

(22)

By means other than Ward's method, there are the cases to obtain the reduced distances after merging clusters by median point. That is to say, it cannot ensure the monotonicity of distance. Therefore, we will use Ward method with measuring the inter-cluster.

### 3.4. Visualization System of Purchase and Position Information

We were implementation the visualization system based on the above methods (section3.3) as shown in the Fig7. "Purchase Information Network Map of Shinjuku-City" in the Fig7 is the example of Purchase Information Network Map. The colour classification of the nodes expresses difference of purchase and position information. This system can visualize the relationship to other Cities, Towns and Villages by this colour classification.

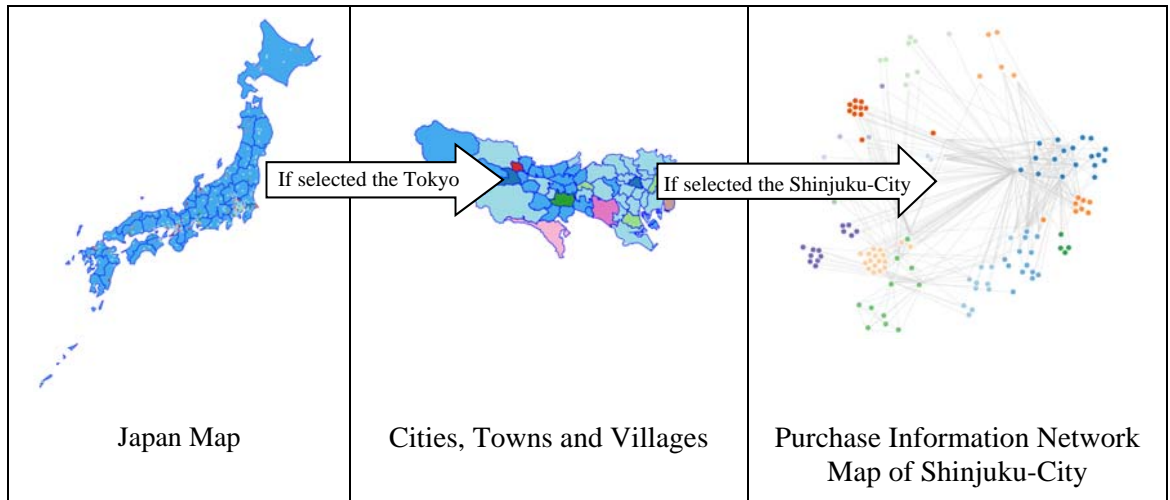

| Japan Map | Cities, Towns and Villages | Purchase Information Network Map of Shinjuku-City |

Figure 7. Visualization System of Purchase and Position Information

## 4. EVALUATION EXPERIMENT

### 4.1. Outline of Evaluation Experiment

We will compare the Harada's method ($Q^H$, Section2.4) and our method using actual SNS data (about 700) in this evaluation experiment. We have set the correct community divided ($R_n$) as an index of this evaluation experiment. The n of $R_n$ shows the node, and we had compared the $Q^H$ and our method while increasing increments of 100 nodes. $R_n$ is defined like follows (1. – 4.).

1. If "The information of Purchase _ Purchase place" is the same, these nodes will set at the same community (table3).

2. If "The information of Purchase" is the same, these nodes will set at the same community. It is to be done for the rest nodes of above 1.

3. If "Purchase place" is the same, these nodes will set at the same community. It is to be done for the rest nodes of above 2.

4. Finally, if it does not match with everything node, it will be treated as a single community.

By increasing the number of nodes in the 1 to 4 work by the hundred, changes in cluster numbers are estimated, an estimate called $R_n$.

Fig8 shows the result of $R_n$, QH and our method. As found in Fig8, the proposed method shows a similar pattern to $R_n$. In contrast, until the number of nodes reaches 400, the number of community for $Q^H$ is far larger than that for $R_n$. The $Q^H$ community number, however, is much smaller than that of $R_n$ after the number of nodes exceeds 400.

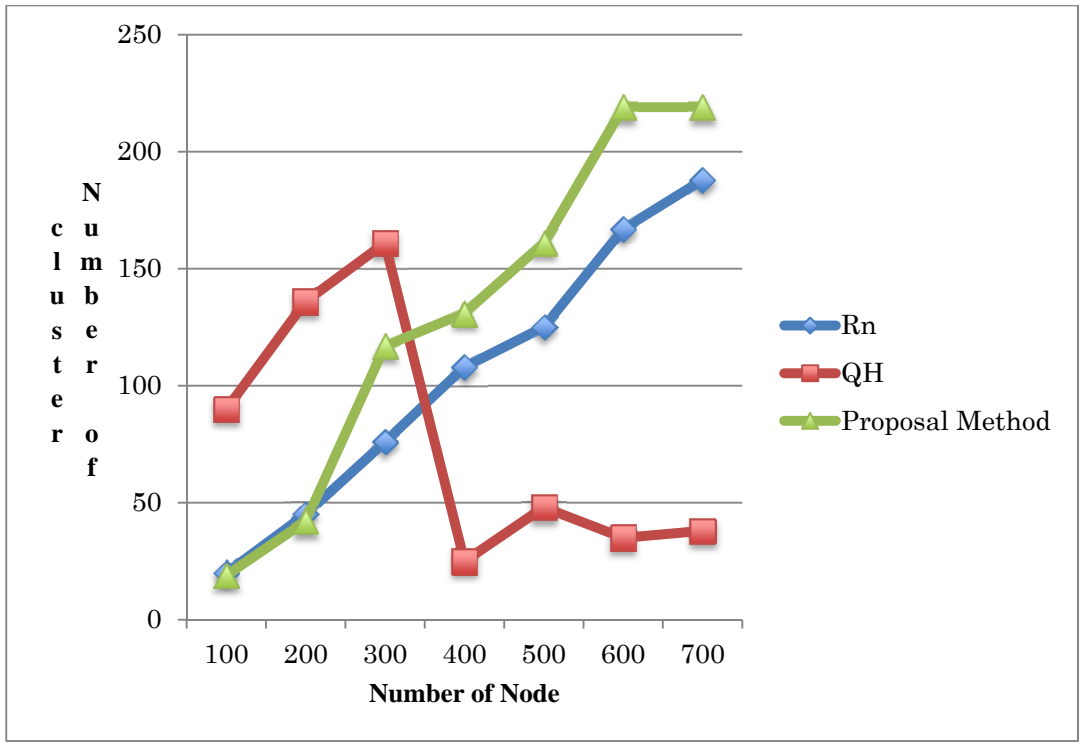

Figure 8. Comparison of community number of $R_n$, $Q^H$ and Proposal Method

Then, A t-test is conducted to determine how much difference there is between the proposed approach and $Q^H$ in terms of the community-number mean value. More specifically, the t-test is on the difference between the mean values of the two data sets when the number of nodes in Fig8 is between 100 and 700. If the null hypothesis is that there is no difference between the mean values of the proposed approach and $Q^H$, the t-test result is: 0.046<0.05, the level of significance. It has thus been found that there is a statistically significant difference between the two.

### 4.2. Discussion of Evaluation Experiment

This section will discuss the above evaluation experiment. If communities are created based on social networking purchase data, as shown in the $R_n$ definition, the number of communities are very likely to increase according to the number of nodes. The reason for this is because the corresponding number itself decreasing is not possible, since the corresponding number among nodes would accumulate proportionally to the increase of nodes.

The $Q^H$ community number, however, is much smaller than that of $R_n$ after the number of nodes exceeds 400. The cause of this is considered as follows. $Q^H$ is realized in the form of expanding modularity $Q$. Modularity $Q$ is something that carries out division in a condition where the connections within the same community are at the most dense yet the connections to other communities are at the least. In other words, since modularity has the characteristics of becoming more coherent as the number of similar references increases, this result in the division being carried out with smaller number of communities. This trait appears to have become prominent in this experiment after the number of nodes exceeded 400. However, as the proposed technique takes into account this correspondence while carrying out clustering, the result has shown transition of community number similar to that of $R_n$.

# 4. CONCLUSION

In this study, Collecting regional information and purchase information from Twitter and representing them as bipartite graph, a technique to analyse "Responsiveness Pair Clustering" has been proposed. The modularity $Q$ can't handle the network if there is restriction on the link. This study was solved this problem by using the "Responsiveness Pair Clustering" instead of the Modularity $Q$. Then we confirmed predominance of our method than $Q^H$ by result of evaluation experiment. Furthermore, we constructed the visualization system of customer purchase based on this method. This visualization system is expected to contribute as an analytical system for customers' purchasing behaviour and so on.

**Authors**

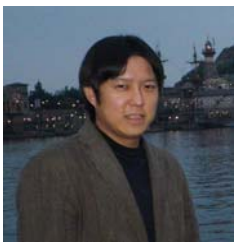

**Akira Otsuki**

Received his Ph.D. in engineering from Keio University (Japan), in 2012. He is currently associate professor at Tokyo institute of technology (Japan) and Officer at Japan society of Information and knowledge (JSIK). His research interests include Analysis of Big Data, Data Mining, Academic Landscape, and new knowledge creation support system. Received his Best paper award 2012 at JSIK. And received his award in Editage Inspired Researcher Grant, in 2012.

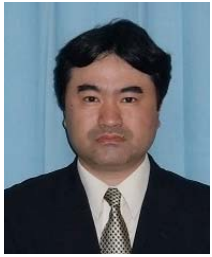

**Masayoshi Kawamura**

Masayoshi Kawamura is a system engineer (Japan). He received M.S. degree from Kyoto Institute of Technology (Japan) in 1998. His research interests include image processing, digital signal processing, and statistical data analysis.